\documentclass[aps,prl,twocolumn,groupaddress,amssymb,amsmath,nofootinbib]{revtex4}

\usepackage{color}
\usepackage{graphicx}
\usepackage{float}
\usepackage[usenames,dvipsnames]{xcolor}

\begin{document}

\preprint{}

\title{Seven largest couplings of the standard model as IR fixed points}

\author{Radovan Derm\' \i\v sek}

\email[]{dermisek@indiana.edu}

\author{Navin McGinnis}

\email[]{nmmcginn@indiana.edu}

\affiliation{Physics Department, Indiana University, Bloomington, IN 47405, USA}


\date{December 13, 2018}

\begin{abstract}

We report on an intriguing  observation
 that the values of all the couplings in the standard model except those related to first two generations can be understood from the IR fixed point structure of renormalization group  equations  in the minimal supersymmetric  model extended by one complete vectorlike family with the scale of new physics in a multi-TeV range.

\end{abstract}


\maketitle






{\bf Introduction.}
Out of seventeen dimensionless couplings of the standard model (SM) only seven are sizable: the three gauge couplings of the $SU(3)\times SU(2)\times U(1)$ symmetry, the top, bottom and tau Yukawa couplings leading to masses of third generation fermions and the Higgs quartic coupling inferred from the Higgs boson mass. The next largest coupling, the Yukawa coupling of the charm quark, is less than 1\% of the top Yukawa coupling. 

The origin of the SM parameters is an open question. Larger or additional symmetries at a high scale can provide relations between some of the couplings as, for example, in grand unified theories (GUT) or models with family symmetries. Alternatively, the structure of the renormalization group (RG) equations of a given model may lead to a certain pattern in model parameters far below the fundamental scale that depends very little  on boundary conditions. Indeed, there have been many attempts to understand values of some gauge couplings \cite{Maiani:1977cg, Cabibbo:1982hy, Moroi:1993, Dermisek:2012as, Dermisek:2012ke, Dermisek:2017ihj} or Yukawa couplings \cite{Pendleton:1980as, Hill:1980sq, Bardeen:1993rv, Carena:1993bs, Lanzagorta:1995gp, Bando:1997dg, Ghilencea:1997yr, Dermisek:2018hxq} from the IR fixed point structure of RG equations. This is quite an intriguing possibility that allows for predictions of SM parameters  even if the fundamental symmetries or model parameters at the fundamental scale remain obscure.

We show that all the couplings of the SM except those related to first two generations can be understood from the IR fixed point structure of RG equations  in the minimal supersymmetric   model extended by one complete vectorlike family (MSSM+1VF) with the scale of new physics in a multi-TeV range. The pattern of seven largest couplings can be predicted in terms of three parameters:  
\begin{equation}
M_G, \; M, \; \tan \beta,  
\label{eq:mass_pars}
\end{equation}
representing the GUT scale (or the fundamental scale of the model, since we do not necessarily assume grand unification), the scale of new physics (vectorlike matter and superpartners, $M \equiv M_{V} = M_{SUSY}$),  and the ratio of vacuum expectation values of the two Higgs doublets, $\tan \beta = v_u/v_d$. Random large values of couplings at  $M_G$ inevitably lead to electroweak (EW) scale values very close to the observed ones.

 \begin{figure}[b]
\centering
\vspace{-1.3cm}
\includegraphics[width = 3.in]{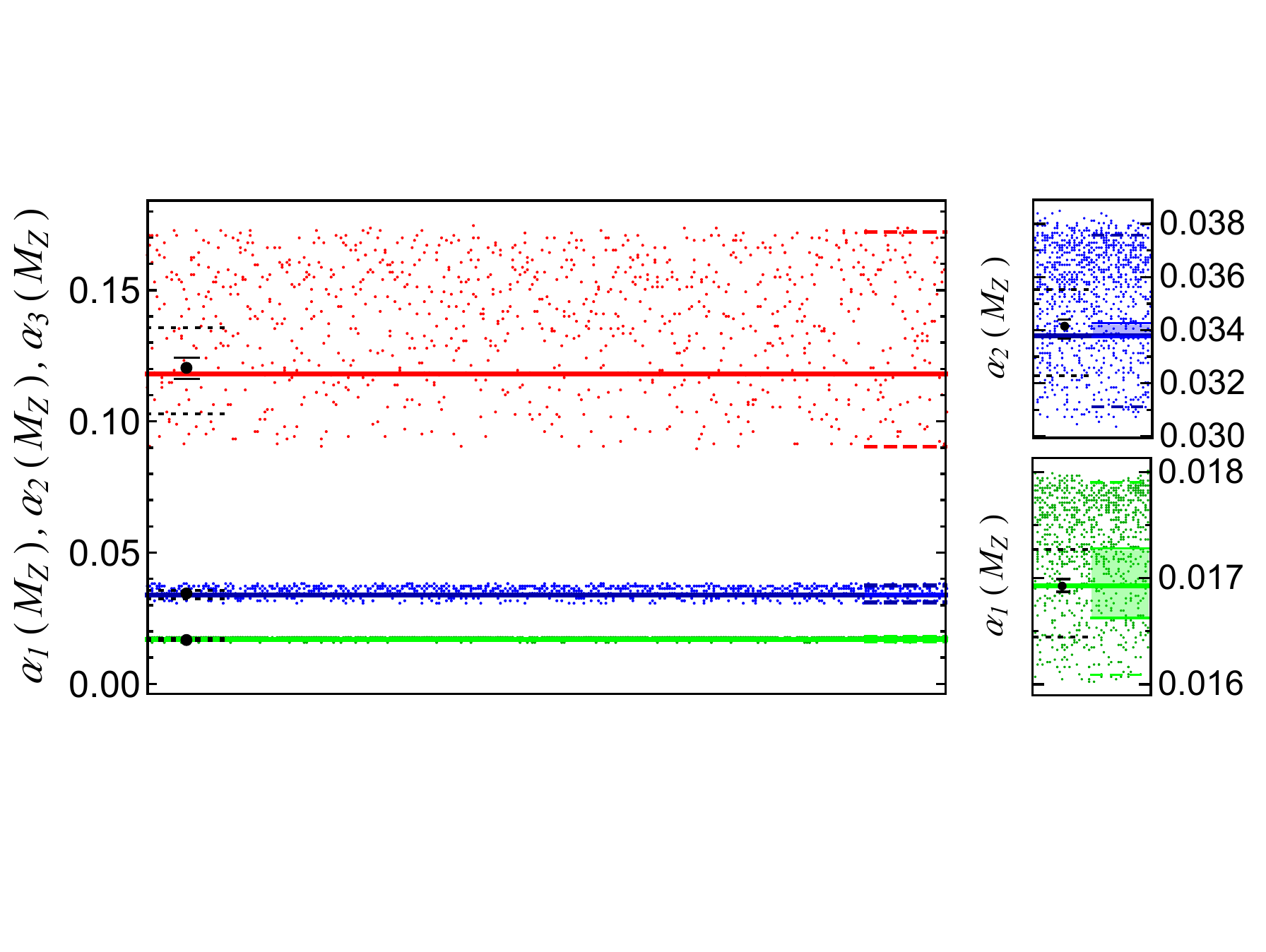}\\
\vspace{-2.2cm}
\includegraphics[width = 2.97in]{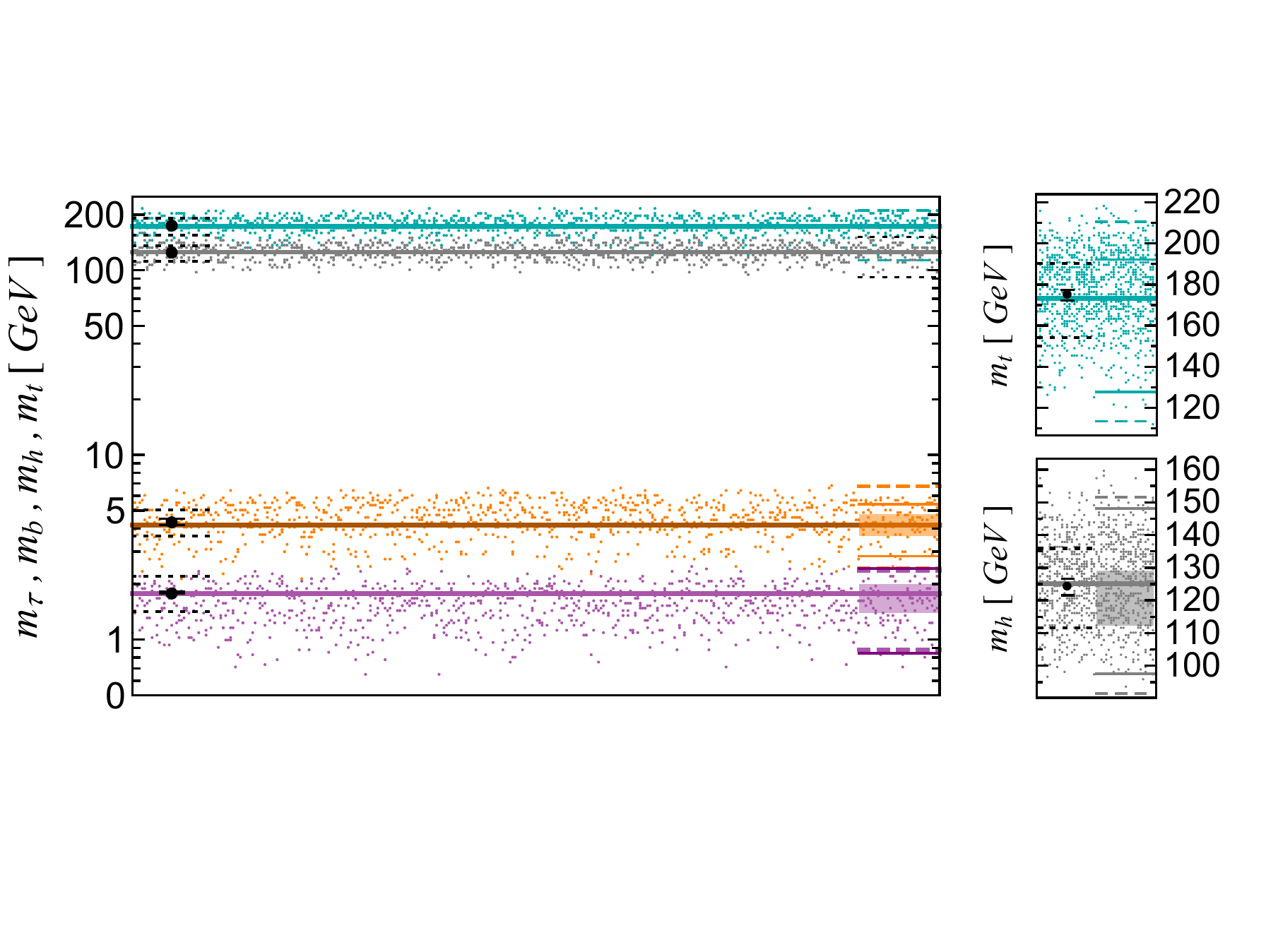}\\
\vspace{-1.3cm}
\caption{Predicted values of $\alpha_{1}$, $\alpha_2$, $\alpha_3$ at $M_Z$, $m_t$, $m_b$, $m_\tau$ and $m_h$ for randomly generated $\alpha_1(M_G)$, $\alpha_2(M_G)$, $\alpha_3(M_G)\in [0.1,0.3]$ and $y_t(M_G)$, $y_b(M_G) $, $y_\tau(M_G)$, $Y_V \in [1,3]$  with fixed $M_G = 3.5\times 10^{16} $~GeV, $M = 7$~TeV and $\tan \beta = 40$. Solid lines indicate the measured central values. Dashed lines at the right edges  indicate the span of predictions assuming universal GUT scale boundary conditions for gauge and Yukawa couplings in the same ranges. The thin solid lines at the right edges further assume that $M$ is optimized to fit the central measured value of $\alpha_3$  and the shaded regions further assume that $Y_V$ is optimized to fit the central measured value of $m_t$. The highlighted point corresponds to the scenario in Fig.~\ref{fig:RG}, the small solid lines near the point correspond to varying $M$ by $\pm 20\%$ and the dashed lines at the left edges  indicate the span of predictions when all couplings are varied in $\pm20\%$ ranges around the values corresponding to the highlighted point. Selected couplings and masses are also zoomed in.}
\label{fig:random}
\end{figure}

The main results are summarized in Fig.~\ref{fig:random} where we plot the predicted EW scale values of the gauge couplings, the third generation fermion masses and the Higgs boson mass for randomly generated and uncorrelated boundary conditions of gauge and Yukawa couplings in large intervals at the GUT scale (we plot the masses instead of couplings because these are more recognizable).
We see that the predictions from random boundary conditions closely cluster around  the measured central values (solid lines). 
Almost identical results are obtained assuming universal gauge and Yukawa couplings at $M_G$ in the same intervals (indicated by dashed lines at the right edges). This clearly points to the IR fixed point nature of predictions at the EW scale.  

Furthermore, if the scale of new physics is optimized to fit the measured central value of $\alpha_3$ and if universal Yukawa couplings are optimized to fit the central measured value of the top quark mass the span of predictions for other gauge couplings, Yukawa couplings and the Higgs quartic coupling significantly shrinks (thin solid lines and shaded regions at the right edges). 

\begin{figure}[t]
\centering
\includegraphics[width = 3.in]{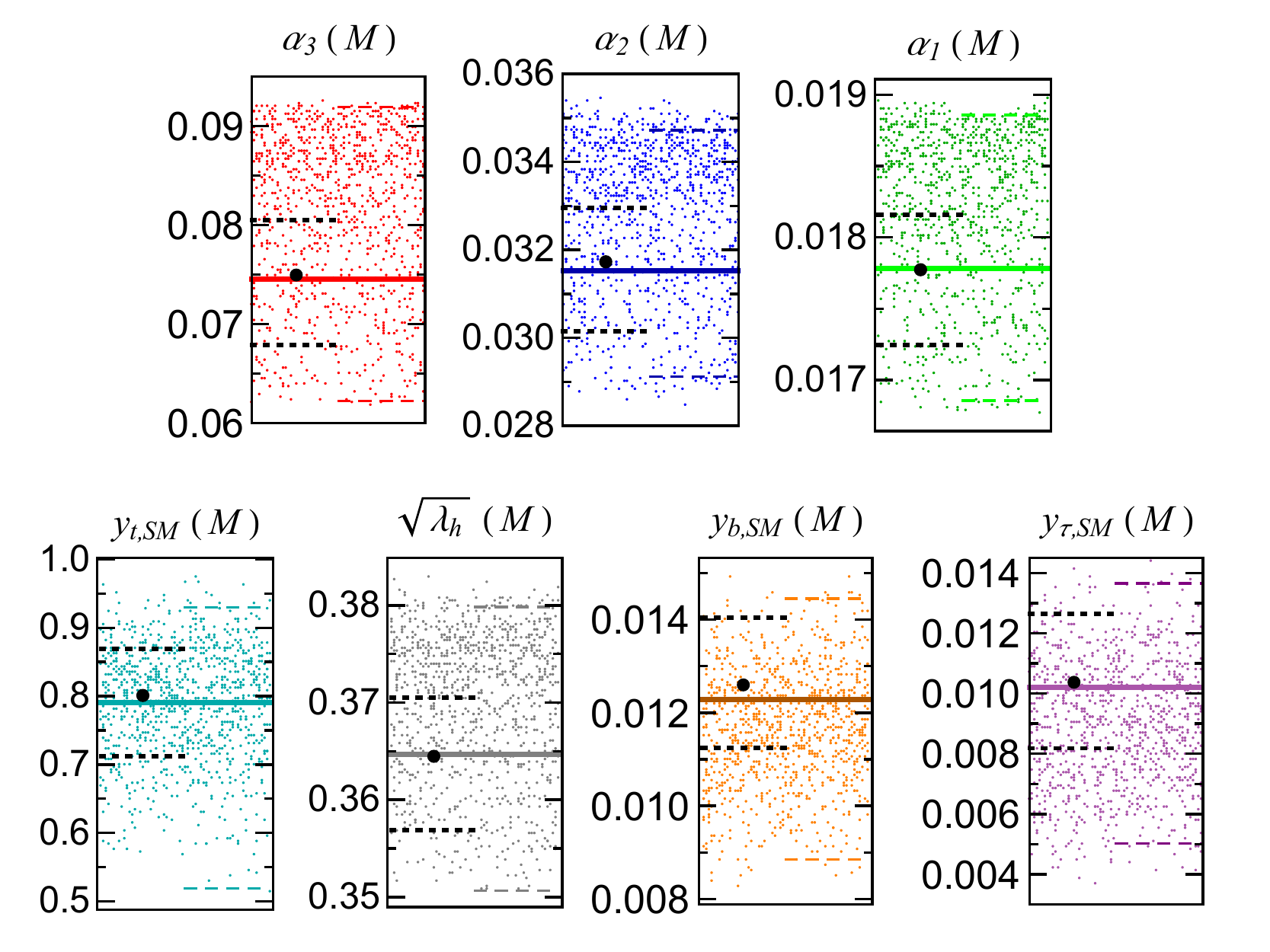}
\caption{Points, lines and the highlighted point represent the same as in Fig.~\ref{fig:random} but with couplings evaluated at $M = 7$~TeV.  
}
\label{fig:random_M}
\end{figure}

More 
importantly, the IR fixed point predictions for couplings in the MSSM+1VF  should be compared with the SM values at the scale of new physics. We plot the corresponding information in Fig.~\ref{fig:random_M}.  We see that the relative ranges for $\alpha_3$, $\lambda_h$ and $y_b$ at $M$ are significantly smaller and the predictions are sharper. This is due to the fact that these couplings change the most in the RG evolution below the scale of new physics.

   \begin{figure}[t]
\centering
\includegraphics[width = 2.8in]{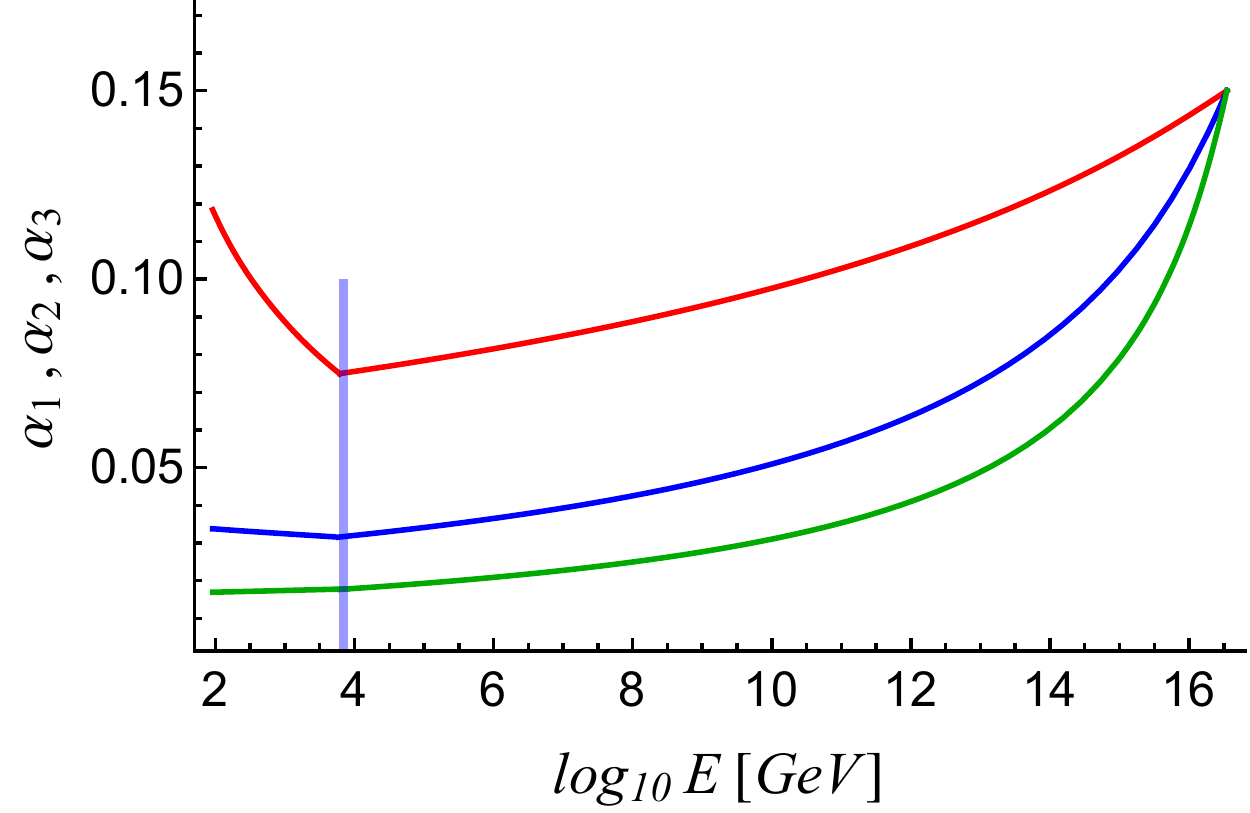}\\
\includegraphics[width = 2.7in]{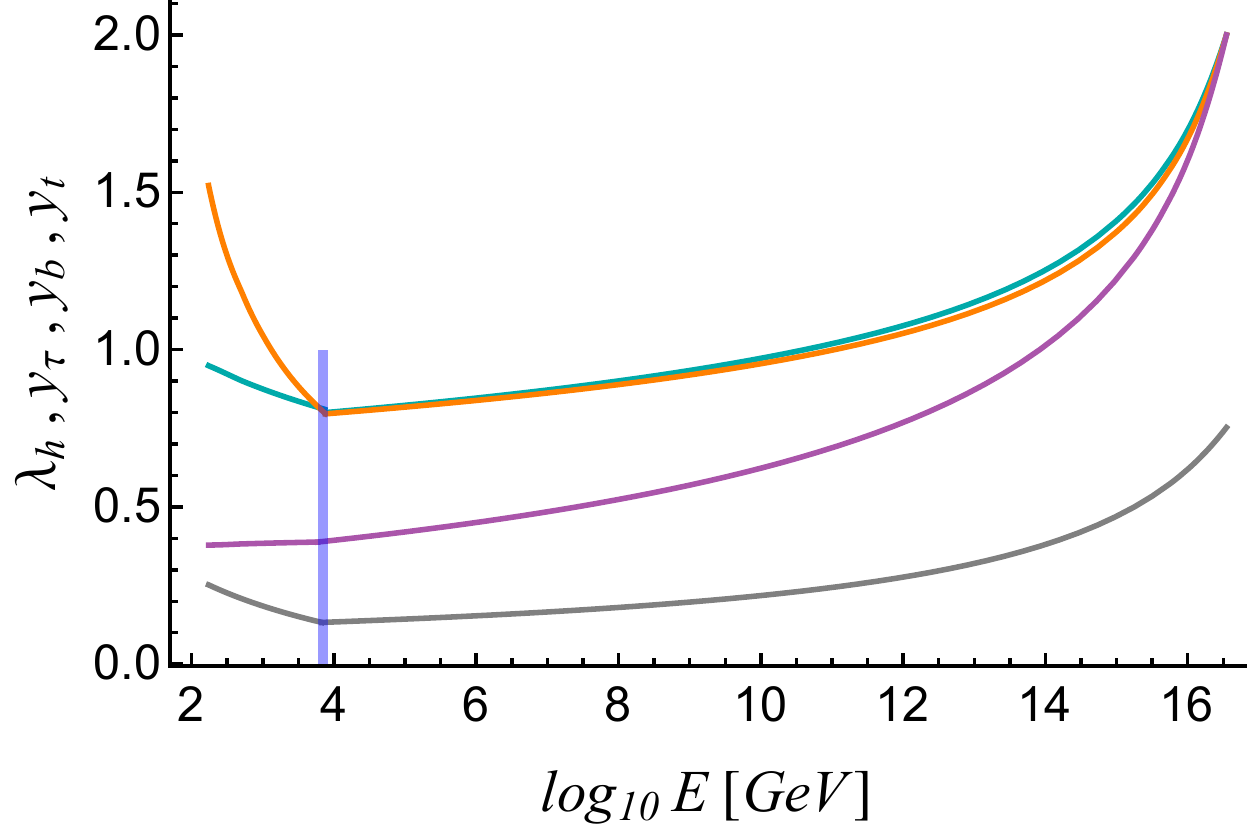}
\caption{RG evolution of $\alpha_{1}$, $\alpha_2$, $\alpha_3$, $y_t$, $y_b$, $y_\tau$ and $\lambda_h$ (right-hand side of Eq.~(\ref{eq:lambda_h})) in the MSSM+1VF starting from $\alpha_G = 0.15$ and $Y_0 =2$ at $M_G = 3.5 \times 10^{16} $~GeV, assuming $Y_V=1.6$.  At low energies, the evolution of  $\alpha_{1,2,3}$ and $\lambda_h$ in the SM starts from the measured central values.  The plotted $y_{t,b,\tau}$ at low energies are obtained from Eqs.~(\ref{eq:yt-corrections}) -- (\ref{eq:ytau-corrections}) and the measured central values of fermion masses, assuming  $\tan\beta =40$  and that all superpartners and Higgs bosons except the SM-like one are at the corresponding RG scale,  $M=E$. The vertical highlight shows the range of $M$, approximately  6 TeV - 8 TeV, within which all the  couplings evolved using  SM RG equations meet the corresponding couplings in the MSSM+1VF.} 
\label{fig:RG}
\end{figure}

  The RG evolution for one set of boundary conditions is given in Fig.~\ref{fig:RG}. In this example (for simplicity chosen with universal boundary conditions at $M_G$),  all the couplings in the MSSM+1VF meet the corresponding parameters in the SM within a very narrow range of $M$, approximately  6 TeV - 8 TeV. 
 Assuming comparable boundary conditions, the RG evolution of individual vectorlike Yukawa couplings closely follows the evolution of $y_t$ or $y_b$ (for quark Yukawas) and $y_\tau$ (for lepton Yukawas) reaching almost identical values at low energies  and thus we do not show it. 

The example from Fig.~\ref{fig:RG} is shown as a highlighted point in Figs.~\ref{fig:random} and \ref{fig:random_M}. We also show  the span of predictions when all couplings are independently varied in $\pm20\%$ ranges around the values corresponding to the highlighted point (dashed lines on the left) and in Fig.~\ref{fig:random}  we also indicate the impact of varying $M$ by $\pm 20\%$ (small solid lines near the point). Varying $M_G$ in $\pm 20\%$ range has a much smaller impact and varying $\tan\beta$ only affects the overall scale of bottom and $\tau$ masses (in the large $\tan \beta$ region even significant variations have a negligible impact on the top quark mass). 
Before we discuss the behavior and predictions for individual couplings further, we summarize the model and details of the analysis.


{\bf Model and details of the analysis.}
The MSSM+1VF is the minimal supersymmetric model extended by a complete vectorlike family (an exact copy of a SM family: $q$, $\bar u$, $\bar d$, $l$, $\bar e$ and corresponding fields with conjugate quantum numbers). The three parameters in Eq.~(\ref{eq:mass_pars}), related to scales of the model,
are the most important parameters determining the values of dimensionless couplings in the SM. The results do not differ much if the common mass of superpartners, $M_{SUSY}$, is not identified with  the common mass of vectorlike fermions, $M_V$, as long as these two scales are comparable. 
 Similarly, 
assuming a split spectrum of vectorlike matter or superpartners, for example the spectrum
obtained from the RG evolution starting with common mass terms at the GUT scale, would only result in logarithmic threshold corrections to the gauge and third generation Yukawa couplings. Unless the splitting is huge these effects are negligible compared to the span of outcomes from the ranges of boundary conditions we consider.
Thus, for simplicity, we assume only one scale of new physics, $M$, at which the SM is matched to the MSSM+1VF.

The RG evolution of dimensionless parameters starts at $M_G$. At this scale, we consider unrelated boundary conditions for gauge couplings, $\alpha_{1,2,3} (M_G) \in [0.1,0.3]$,  and unrelated boundary conditions for  the third generation Yukawa couplings, $y_{t,b,\tau} (M_G) \in [1,3]$.
In  a subset of results, we also consider universal boundary conditions, $\alpha_G(M_G)$ and $Y_0(M_G)$,  in the same intervals. 
For simplicity and also not to favor contributions to the top, bottom or $\tau$ Yukawa couplings in the RG evolution,  we assume a common Yukawa coupling for all  vectorlike fields, $Y_V(M_G)$ (assuming unrelated boundary conditions for individual Yukawa couplings of vectorlike fields in the same interval does not lead to a visible difference in plotted distributions).
We also neglect Yukawa couplings of the first two SM generations and possible mixing between the third generation and vectorlike matter.

Below the GUT scale, 
we use 3-loop RG equations for gauge couplings and 2-loop RG equations for the third generation Yukawa couplings and Yukawa couplings of vectorlike fields based on Refs.~\cite{Jones:1975, Machacek:1983tz, Machacek:1983fi, Machacek:1984zw, Martin:1993, Castano:1993ri, Kolda:1996ea}.  The RG equations  for  gauge  and Yukawa couplings in this model and related discussion can be found in Refs.~\cite{Dermisek:2017ihj, Dermisek:2018hxq}, where predictions for the scale of new physics from gauge coupling unification and the possibility of Yukawa coupling unification in this model were studied. 

All the particles above the EW scale are integrated out at their corresponding mass scales. The complete set of SUSY threshold corrections to the third generation Yukawa couplings is included at the $M_{SUSY}$ scale, identified with $M$, where the  Yukawa couplings in the SM are generated:
\begin{eqnarray}
y_{t,SM} (M) &=& y_t (M) \sin \beta \, (1 + \epsilon_t), \label{eq:yt-corrections} \\
y_{b,SM} (M) &=& y_b (M) \cos \beta \, (1 + \epsilon_b),\\
y_{\tau,SM} (M) &=& y_\tau (M) \cos \beta \, (1 + \epsilon_\tau) \label{eq:ytau-corrections},
\end{eqnarray}
with $\epsilon_{t,b,\tau}$ representing SUSY threshold corrections~\cite{Hall:1993gn, Hempfling:1993kv, Carena:1994bv}. The corrections assume zero soft trilinear couplings (A-terms) and  the supersymmetric Higgs mass $\mu = - \sqrt{2} M_{SUSY}$ (the absolute value is motivated by radiative EW symmetry breaking and the sign is favored by the measured bottom quark mass). The approximate formulas for the corrections with these assumptions can be found in Ref.~\cite{Dermisek:2018hxq}. 

Similarly, the Higgs quartic coupling is generated at the scale of new physics:
\begin{equation}
\lambda_h ( M) \; \equiv \; \frac{g_2^2 (M) + (3/5) g_1^2 (M)}{4}  \; \cos^2 2 \beta. 
\label{eq:lambda_h}
\end{equation}
We assume no finite threshold corrections at the $M$ scale (consistent with  zero A-terms that we assumed for threshold corrections of Yukawa couplings). Below this scale, $\lambda_h$ is evolved using  the SM 2-loop RG equation.

For the measured central values of gauge couplings, fermion masses and the Higgs boson mass we use:  $\alpha_{EM}^{-1} (M_Z) = 127.955$, $\sin^2 \theta_W = 0.2312$, $\alpha_{3} (M_Z) = 0.1181$, $m_t = 173.1$ GeV, $m_b(m_b) = 4.18$ GeV, $m_\tau = 1.777$ GeV, where $m_t$ and $m_\tau$ are pole masses, and $m_h = 125.2$ GeV~\cite{Tanabashi:2018oca}. From these we obtain the corresponding running couplings~\cite{Pierce:1997}.


{\bf Discussion of results.}
The behavior of gauge couplings in the MSSM+1VF can be qualitatively understood  from the solution of the one-loop RG equations,
\begin{equation}
\alpha_i^{-1}(M)=\frac{b_i}{2\pi}\ln\frac{M_G}{M} + \alpha_i^{-1}(M_G),
\label{eq:a}
\end{equation}
where $b_i = (53/5,5,1)$. Note that adding one complete vectorlike family makes all three couplings asymptotically divergent. Thus, starting with large (but still perturbative) boundary conditions  at $M_G$, in the RG evolution to low energies, the  gauge couplings run to the (trivial) IR fixed  point. Far below the GUT scale, the log term dominates and the couplings approach zero at fixed ratios given by the ratios of beta function coefficients (that the ratios of gauge couplings are approximately constant  far away from the GUT scale is visible in Fig.~\ref{fig:RG}). The boundary conditions (and thus whether gauge couplings unify or not) are less and less important  further away from the GUT scale we evolve. 

From Figs.~\ref{fig:random} and \ref{fig:random_M} we see that the IR fixed point behavior is very effective for $\alpha_1$ and $\alpha_2$ with all the points  within about $\pm5\%$ and $\pm10\%$ from the central values. As a result of the small beta function coefficient $\alpha_3$ approaches the trivial fixed point at much slower pace and thus depends on the boundary condition the most. 
The span of $\alpha_3(M)$ is  about $\pm 15\%$ around the central value. Because $\alpha_3$ runs the fastest in the SM, it is the most determining factor for the scale of new physics. We can  fix $M$ to fit the measured central value of $\alpha_3$. For the range of the scan,  $\alpha_3(M_G)\in [0.1,0.3]$, we find $M \in$ [1 TeV, 45 TeV] with the lower bound motivating the upper bound of $\alpha_3(M_G)$ in the scans. 
Fixing $M$ to fit $\alpha_3$ inevitably leads to $\alpha_1$ and $\alpha_2$ being very close to measured values since predictions for these depend very little on $M$ and even less on the boundary conditions at $M_G$.\footnote{The agreement is especially striking assuming random universal boundary conditions at $M_G$, shown by shaded ranges in Figs.~\ref{fig:random}.
Predictions from gauge coupling unification agree with measured values even better than in the MSSM as a result of 2-loop effects from large Yukawa couplings.}

The quark Yukawa couplings (top, bottom and those of vectorlike fields) approach the IR fixed point determined mostly  by $\alpha_3$ very fast. The IR fixed point behavior is extremely effective.  All large quark Yukawa couplings share the IR fixed point value and thus the number of large quark Yukawa couplings is more important than their boundary conditions. The IR fixed point values depending on the number of vectorlike quark Yukawa couplings can be calculated assuming common boundary conditions~\cite{Dermisek:2018hxq} and these results remain a good approximation as long as the boundary conditions are comparable.
Lepton Yukawa couplings ($\tau$ and those of vectorlike fields) are driven to the trivial fixed point by large quark Yukawa couplings (and thus indirectly by $\alpha_3$). 
From  Figs.~\ref{fig:random} and \ref{fig:random_M} we see that 
predictions from large range of uncorrelated  boundary conditions  cluster around the measured central values. Although the predictions span larger ranges than those of gauge couplings, we should keep in mind that we are scanning a huge range of possible independent boundary conditions that 
 span almost an order of magnitude (1/3 - 3) of their ratios  (this affects especially $m_\tau$). As the shaded ranges in Fig.~\ref{fig:random} show, assuming random universal boundary conditions and optimizing $Y_V$ to fit the measured central value of $m_t$ (that requires comparable $Y_V$ and $Y_0$) leads to much smaller ranges of predicted values of $m_b$ and $m_\tau$ that closely cluster around the measured values.

Finally, since $\lambda_h(M)$ is given in terms of $\alpha_1$ and $\alpha_2$ and these are sharply focused far below the GUT scale, the predicted range of $\lambda_h(M)$ is also very narrow. The much larger spread of predictions at the $M_Z$ scale is due to very fast running in the SM between $M$ and $M_Z$. However, at the $M$ scale the IR fixed point prediction in the MSSM+1VF is very sharp, see Fig.~\ref{fig:random_M}. 
The agreement with the measured value follows from the fact that $m_h \simeq 125$ GeV, in the absence of mixing in the stop sector, requires ${\cal O}(10 \;{\rm TeV})$ stop masses~\cite{Hahn:2013ria, Draper:2013oza}.


{\bf Conclusions.}
We have shown that the seven largest couplings of the SM can be predicted in terms of three parameters 
related to mass scales from a huge range of random order one\footnote{Note that the lower limit for gauge couplings we consider, $\alpha_i = 0.1$, implies lagrangian parameters $g_i \simeq1$ since $\alpha_i = g_i^2/4\pi$.} or larger  couplings at the GUT scale as long as the masses of vectorlike family and superpartners are in a multi-TeV range.
Precise predictions of course require precise input values of all the couplings. Nevertheless, due to the IR fixed point structure of RG equations, rough predictions can be made without detailed knowledge of boundary conditions.

For the ranges of unrelated (or unified) boundary conditions that we considered, spanning a factor of three between the largest and smallest, we find that the  parameters in Eq.~(\ref{eq:mass_pars}) can be optimized so that none of the seven observables is more than 25\%  (or 15\%) from the measured values. Further optimizing $Y_V$ to obtain the required overall scale of Yukawa couplings we find all seven observables within 11\%  (or 7.5\%) from their measured values. Predictions are even sharper at the scale of new physics.

This scenario takes advantage of the  success of gauge coupling unification in the MSSM and 
the prediction for the Higgs boson mass (or $\lambda_h$) in supersymmetric theories. However, that all seven couplings can be understood simultaneously from random large boundary conditions at the GUT scale  assuming only  one (and common) scale of new physics is highly non-trivial. Repeating a similar exercise in the MSSM, randomly selecting gauge and Yukawa couplings in ranges $\pm50\%$  around the best fit values (that corresponds to the factor of 3 between the largest and the smallest as in our scans), no familiar pattern would emerge. Predicted $\alpha_3(M_Z)$ would span  an order of magnitude with no gaps between three gauge couplings. Predictions would include $\alpha_3$ comparable to $\alpha_1$ with much larger $\alpha_2$, $m_t$ below $M_Z$ and above  400 GeV, over an order of magnitude range for $m_b$ and the $\tau$ often much heavier than the bottom quark.

Our findings motivate  new  fermions and superpartners with masses in a multi-TeV range. Although the typical scale is beyond the reach of the LHC, a part of the spectrum could be within the reach if $M_V$ is split from $M_{SUSY}$ or when non-universal vectorlike masses or superpartner masses are assumed. In addition, assuming non-zero A-terms  lowers the scale of new physics required by the Higgs boson mass. Even a significant splitting of masses or introduction of moderate A-terms does not affect the presented results significantly, it only moves the  preferred range of couplings (smaller scales of new physics prefer larger gauge and Yukawa couplings at $M_G$).

Furthermore, since in this model the fundamental symmetries or relations between couplings at the fundamental scale are not crucial for predictions that would agree with observations, this may open a new direction for exploring other UV embeddings of the SM that might shed light on different features of the SM or remove problematic aspects of conventional GUT models (origin of three families, hierarchies in fermion masses, doublet-triplet splitting and proton decay among others).  For example, product group embeddings of the SM, like Pati-Salam or flipped SU(5) among many others, that do not feature gauge or Yukawa coupling unification, would have essentially the same predictions at low energies as models with unification, if extended by a complete vectorlike family.


\vspace{0.2cm}
\noindent
{\bf Acknowledgments:} This work was supported in part by the U.S. Department of Energy under grant number {DE}-SC0010120.




\begin{thebibliography}{99}






\bibitem{Maiani:1977cg} 
  L.~Maiani, G.~Parisi and R.~Petronzio,
  Nucl.\ Phys.\ B {\bf 136}, 115 (1978).

\bibitem{Cabibbo:1982hy} 
  N.~Cabibbo and G.~R.~Farrar,
  Phys.\ Lett.\ B {\bf 110}, 107 (1982).
  
    
\bibitem{Moroi:1993} 
 T.~Moroi, H.~Murayama, and T.~Yanagida,
  Phys.\ Rev.\ D {\bf 48}, 2995 (1993)
  [hep-ph/9306268].
  
  
\bibitem{Dermisek:2012as} 
  R.~Dermisek,
  Phys.\ Lett.\ B {\bf 713}, 469 (2012)
  [arXiv:1204.6533 [hep-ph]].
 
\bibitem{Dermisek:2012ke} 
  R.~Dermisek,
  Phys.\ Rev.\ D {\bf 87}, no. 5, 055008 (2013)
  [arXiv:1212.3035 [hep-ph]].
  
  
  
\bibitem{Dermisek:2017ihj} 
  R.~Dermisek and N.~McGinnis,
  Phys.\ Rev.\ D {\bf 97}, no. 5, 055009 (2018)
  [arXiv:1712.03527 [hep-ph]].
  
  



 
   
   
\bibitem{Pendleton:1980as} 
  B.~Pendleton and G.~G.~Ross,
  Phys.\ Lett.\  {\bf 98B}, 291 (1981).
  
  
\bibitem{Hill:1980sq} 
  C.~T.~Hill,
  Phys.\ Rev.\ D {\bf 24}, 691 (1981).
  

  
  
\bibitem{Bardeen:1993rv} 
  W.~A.~Bardeen, M.~Carena, S.~Pokorski and C.~E.~M.~Wagner,
  Phys.\ Lett.\ B {\bf 320}, 110 (1994)
  [hep-ph/9309293].
  
  
\bibitem{Carena:1993bs}
  M.~Carena, M.~Olechowski, S.~Pokorski and C.~E.~M.~Wagner,
  Nucl.\ Phys.\ B {\bf 419} (1994) 213
  [hep-ph/9311222].


\bibitem{Lanzagorta:1995gp} 
  M.~Lanzagorta and G.~G.~Ross,
  Phys.\ Lett.\ B {\bf 349}, 319 (1995)
  [hep-ph/9501394].


\bibitem{Bando:1997dg} 
  M.~Bando, J.~Sato and K.~Yoshioka,
  Prog.\ Theor.\ Phys.\  {\bf 98}, 169 (1997)
  [hep-ph/9703321].
  
  
\bibitem{Ghilencea:1997yr} 
  D.~Ghilencea, M.~Lanzagorta and G.~G.~Ross,
  Phys.\ Lett.\ B {\bf 415}, 253 (1997)
  [hep-ph/9707462].
  


\bibitem{Dermisek:2018hxq} 
  R.~Dermisek and N.~McGinnis,
  Phys.\ Rev.\ D {\bf 99}, no. 3, 035033 (2019)
  [arXiv:1810.12474 [hep-ph]].


  
  
  
  



  \bibitem{Jones:1975}
  D.R.T.~Jones,
  Nucl.\ Phys. \ B {\bf 87} 127 (1975).

\bibitem{Machacek:1983tz} 
  M.~E.~Machacek and M.~T.~Vaughn,
  Nucl.\ Phys.\ B {\bf 222}, 83 (1983).
 %
\bibitem{Machacek:1983fi} 
  M.~E.~Machacek and M.~T.~Vaughn,
  Nucl.\ Phys.\ B {\bf 236}, 221 (1984).
%
\bibitem{Machacek:1984zw} 
  M.~E.~Machacek and M.~T.~Vaughn,
  Nucl.\ Phys.\ B {\bf 249}, 70 (1985).
  
\bibitem{Martin:1993}
S.~P.~Martin and M.~T.~Vaughn,
Phys.\ Rev.\ D {\bf 50}, 2282 (1994)
[hep-ph/9311340]

  
\bibitem{Castano:1993ri} 
  D.~J.~Castano, E.~J.~Piard and P.~Ramond,
  Phys.\ Rev.\ D {\bf 49}, 4882 (1994)
  [hep-ph/9308335].
  
  
\bibitem{Kolda:1996ea} 
  C.~F.~Kolda and J.~March-Russell,
  Phys.\ Rev.\ D {\bf 55}, 4252 (1997)
  [hep-ph/9609480].




  
\bibitem{Hall:1993gn} 
  L.~J.~Hall, R.~Rattazzi and U.~Sarid,
  Phys.\ Rev.\ D {\bf 50}, 7048 (1994)
  [hep-ph/9306309].
  
\bibitem{Hempfling:1993kv} 
  R.~Hempfling,
  Phys.\ Rev.\ D {\bf 49}, 6168 (1994).
  
  
\bibitem{Carena:1994bv} 
  M.~Carena, M.~Olechowski, S.~Pokorski and C.~E.~M.~Wagner,
  Nucl.\ Phys.\ B {\bf 426}, 269 (1994)
  [hep-ph/9402253].





\bibitem{Tanabashi:2018oca} 
  M.~Tanabashi {\it et al.} [Particle Data Group],
  Phys.\ Rev.\ D {\bf 98}, no. 3, 030001 (2018).
  
  
  

  \bibitem{Pierce:1997}
  D.~M.~Pierce, J.~A.~Bagger, K.~T.~Matchev, R.~Zhang,
  Nucl.\ Phys.\ B {\bf 491}, 1 (1997) 
   [arXiv:hep-ph/9606211].




  
  
\bibitem{Hahn:2013ria} 
  T.~Hahn, S.~Heinemeyer, W.~Hollik, H.~Rzehak and G.~Weiglein,
  Phys.\ Rev.\ Lett.\  {\bf 112}, no. 14, 141801 (2014)
  [arXiv:1312.4937 [hep-ph]].


\bibitem{Draper:2013oza} 
  P.~Draper, G.~Lee and C.~E.~M.~Wagner,
  Phys.\ Rev.\ D {\bf 89}, no. 5, 055023 (2014)
  [arXiv:1312.5743 [hep-ph]].



 

 

\end{thebibliography}
\end{document}